\documentclass[letterpaper,10pt]{article}
\usepackage[margin=1in,footskip=0.25in]{geometry}
\usepackage{graphicx}
\usepackage{epstopdf, epsfig}
\usepackage{subcaption}
\usepackage{amsmath,amssymb}
\usepackage{mathptmx}
\usepackage{stix}
\usepackage[citestyle=authoryear,natbib=true,backend=bibtex]{biblatex} 
\usepackage{authblk}

\author[1]{{Rishabh V. More}\footnote{morer@mit.edu}}
\author[2]{Eugene Pashkovski}
\author[2]{Reid Patterson}
\author[1]{Gareth H. McKinley}
\affil[1]{Department of Mechanical Engineering, Massachusetts Institute of Technology,
Cambridge, MA 02139, USA}
\affil[2]{The Lubrizol Corporation, 29400 Lakeland Blvd., Wickliffe, OH 44092, USA}
\date{}                     
\title{Elasto-Inertial Instability in Torsional Flows of Shear-Thinning Viscoelastic Fluids}

\begin{document}

\maketitle

\begin{abstract}
It is well known that inertia-free shearing flows of a viscoelastic fluid with curved streamlines, such as the torsional flow between a rotating cone and plate, or the flow in a Taylor-Couette geometry, can become unstable to a three-dimensional time-dependent instability at conditions exceeding a critical Weissenberg ($Wi$) number. However, the combined effects of fluid elasticity, shear thinning, and finite inertia (as quantified by the Reynolds number $Re$) on the onset of elasto-inertial instabilities are not fully understood. Using a set of cone-plate geometries, we experimentally explore the entire $Wi$ – $Re$ phase space for a series of non-linear viscoelastic fluids (with rate-dependence quantified using a shear thinning parameter $\beta_P(\dot{\gamma})$). We tune $\beta_P(\dot{\gamma})$ by varying the dissolved polymer concentration in solution. This progressively reduces shear-thinning but leads to finite inertial effects before the onset of elastic instability, thus naturally resulting in elasto-inertial coupling. Time-resolved rheometric measurements and flow visualization experiments allow us to investigate the effects of flow geometry and document the combined effects of varying $Wi$, $Re$, and $\beta_P(\dot{\gamma})$ on the emergence of secondary motions at the onset of instability. The resulting critical state diagram quantitatively depicts the competition between the stabilizing effects of shear thinning and the destabilizing effects of inertia. We extend the curved streamline instability criterion of \cite{pakdel1996elastic} for the onset of purely elastic instability in curvilinear geometries by using scaling arguments to incorporate shear thinning and finite inertial effects. The augmented condition facilitates predictions of the onset of instability over a broader range of flow conditions, thus bridging the gap between purely elastic and elasto-inertial curved streamline instabilities. 
\end{abstract}


\section{Introduction}
Rotational flows with curvilinear streamlines are routinely encountered in many applications, such as transport and handling of fluids, dispensing, spin coating, flows in centrifuges, extruders, flows around objects, lubrication and journal bearing flows, polymer processing, as well as rheometry. The torsional flow between co-axial parallel disks or between a cone and a plate are canonical examples in which a unidirectional shear flow with curved streamlines is generated; hence, they are commonly employed in rheometry and as canonical systems in which to study viscoelastic flow stability. Even in the limit of small cone angles, steady secondary motions and flow instabilities can develop \citep{larson1992instabilities, shaqfeh1996purely} owing to non-linearities stemming from interactions between streamline curvature, fluid inertia, and non-Newtonian fluid properties. 

Specifically, in a cone and plate geometry with cone radius $R$, a cone angle $\theta$ and rotating at speed $\Omega$, a Newtonian fluid with a constant viscosity $\eta_s$ and density $\rho$ experiences the onset of inertially-driven toroidal secondary motions \citep{olagunju1997hopf} and at higher rates, inertial turbulence, when the centrifugal force becomes dominantly larger than the viscous force \citep{sdougos1984secondary}. On the other hand, highly elastic ``Boger fluids" with a constant viscosity $\eta_0$ exhibit a time-dependent transition to an unstable state characterized by an enhanced and fluctuating shear stress (initially interpreted as anti-thixotropic behavior \citep{jackson1984rheometrical}), when sheared beyond a critical value of the dimensionless rotational speed, or Deborah number $De=\tau_s\Omega$, where $\tau_s$ is the (constant for a Boger fluid) shear relaxation time \citep{mckinley1991observations}. The unstable flow manifests as radially propagating logarithmic spiral vortices \citep{oztekin1994quantitative}. 
Furthermore, linear stability analysis of the cone-and-plate flow using different constitutive models reveals the critical role of the dimensionless geometry parameter $1/\theta$ \citep{olagunju1997hopf} on the onset of elastic instabilities, and this sensitivity has also been confirmed by experiments \citep{oztekin1994quantitative, mckinley1995self}. Thus, the instability is a function of both the dimensionless rotational speed, i.e., $De=\tau_s\Omega$ and the dimensionless shear rate $\dot{\gamma} = \Omega/\theta$, i.e., the Weissenberg number $Wi=\tau_s\dot{\gamma}$. A comparison of critical conditions in the $Wi-De$ state diagram with linear stability results indeed validates this codependency \citep{oztekin1994quantitative, mckinley1995self}. Thorough reviews of these earlier studies of viscometric flow instabilities can be found in \cite{larson1992instabilities} and \cite{shaqfeh1996purely}.

\cite{pakdel1996elastic} argued that this type of instability generically arises due to the non-uniform stretching of the polymer molecules in a curvilinear shearing flow, which amplifies the elastic ``hoop stress'' in the fluid. This enters the radial momentum balance and amplifies radial velocity perturbations, making the torsional flow of viscoelastic fluids unstable under sufficiently strong driving conditions. They proposed an instability criterion in terms of the product of $Wi$ and $De$ exceeding a critical magnitude denoted $M_{c}$, and showed that this could be written in the generic form $De\,Wi = \left(\tau_s\Omega \right) \, \left( {\tau_s\dot{\gamma}} \right) > M_{c}^2$. More generally, we can write $De=\tau_s U /\mathcal{R}$, now $\mathcal{R}$ being the (geometry-dependent) characteristic radius of the streamline curvature, and $U$ being the characteristic stream-wise fluid velocity. The ratio $\mathcal{R}/U$ gives the characteristic convective time of a flow experiment. Thus, to take into account, more generally, the curvature of a two-dimensional flow and the action of tensile stress difference along the streamlines, this criterion for the onset of instability can be rewritten as \citep{mckinley1996rheological}:
\begin{equation}\label{eq:eq1}
De\,Wi = \frac{\tau_s \, {U}}{\mathcal{R}} \, \frac{N_1}{\sigma} \geq M_{c}^2, 
\end{equation}
where $U$ is a characteristic streamwise fluid velocity, $\mathcal{R}$ is the geometry-dependent characteristic radius of curvature of the streamline, $N_1$ is the first normal stress difference in the fluid, and $\sigma=\eta_0\dot{\gamma}$ is the shear stress. Furthermore, linear stability analysis of the steady base flow of an Oldroyd-B fluid in a cone and plate geometry \citep{olagunju1993secondary, olagunju1995elastic, olagunju1997hopf}, as well as in a Taylor-Couette geometry \citep{schaefer2018geometric}, shows that the viscosity ratio parameter $\beta_P=\eta_P/\eta_0$ (where $\eta_P$ is the polymer contribution to the fluid viscosity and $\eta_{\infty}$ is the Newtonian plateau viscosity at high shear rates, such that $\eta_0 = \eta_P+\eta_{\infty}$) significantly affects the nature of the elastic instability. The effect of $\beta_P$ can be readily incorporated in the condition for purely elastic instability (Eq.~\ref{eq:eq1}) by substituting the Oldroyd-B result $N_1 = 2\eta_P\tau_s\dot{\gamma}^2$ and $\sigma=(\eta_P+\eta_{\infty})\dot{\gamma}$, which gives
\begin{equation}\label{eq:eq2}
\frac{\tau_s \, {U}}{\mathcal{R}} \, \frac{2\eta_P\tau_s\dot{\gamma}}{\eta_0} = \frac{\tau_s \, {U}}{\mathcal{R}} \, 2\beta_P \tau_s\dot{\gamma} \, \geq \, M_c^2 \implies De\,Wi \,\geq\, \frac{M_{c}^2}{2\beta_P}. 
\end{equation}
In dilute polymer solutions, as $\beta_P \to 0$, the critical shear rate $\dot{\gamma}_c=\Omega_c/\theta$ required for the onset of instability diverges, in accordance with experiments. More recently, \cite{schiamberg2006transitional} have studied the effect of changing the viscosity ratio $\beta_P$ on the onset of secondary motion and the evolution towards a fully-developed non-linear state commonly referred to as `elastic turbulence' in a parallel plate geometry using a polyacrylamide Boger fluid. 

Theoretically, it should be possible to modify the criterion in Eq.~\ref{eq:eq1} for the onset of purely elastic instability in shear-thinning viscoelastic fluids by allowing $\tau_s$ and $\eta_P$ to both be shear-rate dependent and writing them as functions of the applied shear rate $\dot{\gamma}$. 
However, the reduction in the viscosity due to shear thinning means a concomitant increase in inertial effects, which can systematically modify purely elastic instabilities observed in Boger fluids at very low Reynolds numbers $Re \ll 1$. This inherent non-linear coupling between inertia and elasticity in shear-thinning viscoelastic fluids makes the corresponding torsional flows more challenging to understand, especially the critical conditions for the onset of elasto-inertial instabilities. Hence, there have been relatively few studies elucidating the effect of shear thinning on viscoelastic flow stability. \cite{dutcher2013effects}, \cite{schaefer2018geometric}, and \cite{lacassagne2021shear} have investigated shear-thinning–mediated elasto-inertial transitional pathways in Taylor-Couette geometries. Similar observations have also been made in the case of the flow of a shear-thinning viscoelastic fluid through a tube \citep{chandra2019instability}. Each of these studies reveals the inherent coupling of fluid elasticity (which tends to destabilize the base shearing flow) and inertia (which tends to restabilize the unsteady flow of viscoelastic fluids). The resulting stability diagrams are best represented by three-dimensional plots in terms of dimensionless parameters characterizing the geometry, fluid elasticity, and the inertia of the flow \citep{dutcher2009effects}.

Very recently, \cite{datta2022perspectives} reviewed our current understanding of the broader topic of viscoelastic flow instabilities and elastic turbulence. They suggest representing the critical conditions for the onset of viscoelastic flow instabilities in the $Wi-Re$ plane. In this plane, a set of exploratory experiments with a given rate-independent viscoelastic fluid in a fixed geometry trace a line with a slope given by the elasticity number $El=Wi/Re$ (see Fig. 4 in \cite{datta2022perspectives}) eventually intersecting with a corresponding stability boundary demarcating the critical conditions for the onset of instability. Tracing the critical conditions using different viscoelastic fluids and flow geometries enables exploration of the entire phase space, and the various flow states observed will showcase the transitions in the instability mechanisms as inertial, elastic, and geometric effects are varied. In the present work, we investigate the development of time-dependent instabilities in a family of shear-thinning viscoelastic polyisobutylene (PIB) solutions of different PIB concentrations (c$_P$) using a set of cone-and-plate geometries of different cone angles ($\theta$) and radii ($R$) to elucidate the combined effects of fluid elasticity, shear-thinning, inertia, and geometry on the onset of torsional viscoelastic flow instabilities.

\section{Materials and Methods}



\begin{figure}
  \centerline{\includegraphics[width=1\textwidth]{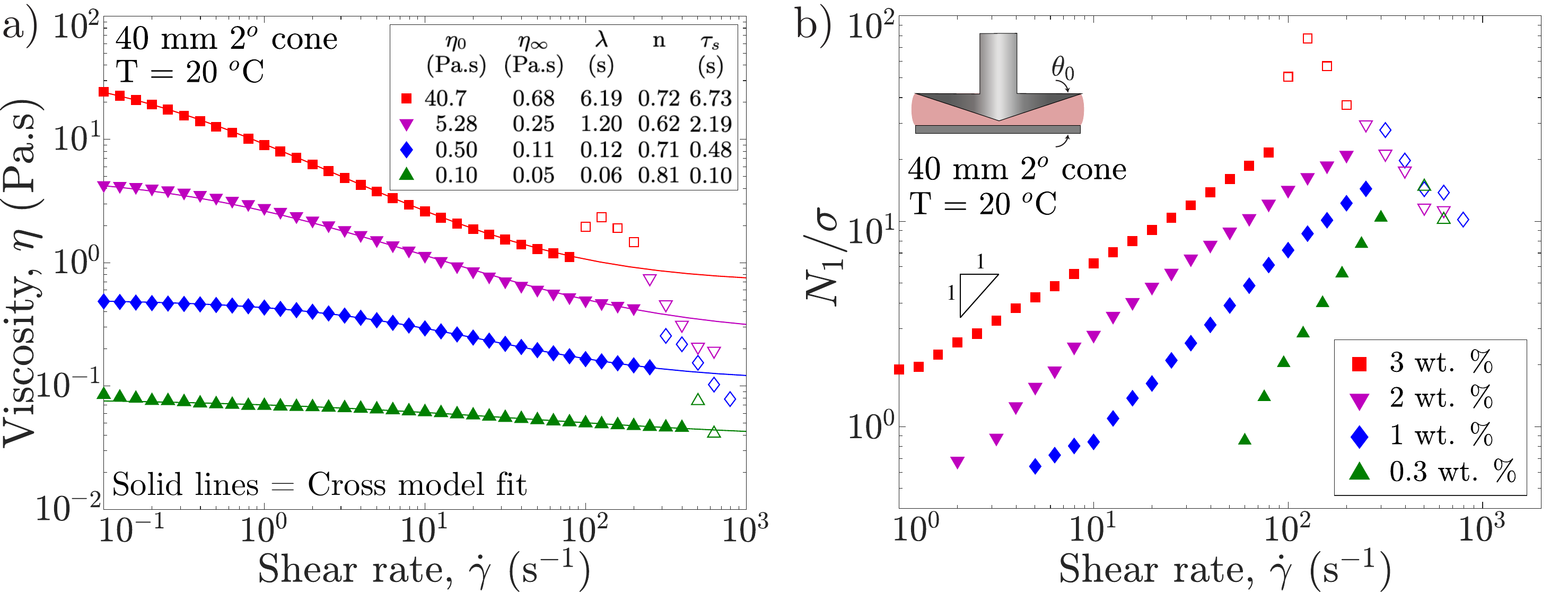}}
  \caption{a) Flow curve and b) rate-dependent stress ratio $S_R=N_1(\dot{\gamma})/\sigma(\dot{\gamma})$ of viscoelastic fluids used in this study. The viscometric properties of the PIB solutions (filled symbols) show a strong shear-thinning behavior, which can be modeled with the Cross model (Solid lines in (a)) before the onset of instability with best-fit parameters presented in the inset of (a). The sudden jump in the viscosity and stress ratio values above a certain critical shear rate (hollow symbols)
  occurs due to the onset of time-dependent flow instabilities arising from a combination of elasticity and inertia.}
\label{fig:fig1-flowcurve}
\end{figure}

We use polymeric solutions of polyisobutylene (PIB) (MW $\approx 10^6$ g/mol) dissolved in a paraffinic oil (GADDTAC, Lubrizol Inc.). We perform all our measurements at a constant temperature $T=20$ $^\circ$C. The Newtonian base oil has a solvent viscosity $\eta_s=18.07$ mPa.s at $20$ $^\circ$C. Dilute solution viscometry determines the polymer intrinsic viscosity to be $\left[ \eta \right] = 3.69$ dL/g, and this gives the critical overlap concentration of the polymer solute as c$^* \simeq 0.77/\left[ \eta \right] = 0.23$ wt.\,\%  \citep{graessley1980polymer}. The solutions were all measured to have a constant density $\rho=873.1$ kg/m$^3$. We vary the dissolved concentration of polymer in the solution to change the viscoelastic properties. We work with three semi-dilute (c$_P$ $>$ c$^*$) solutions: 3 wt.\,\%, 2 wt.\,\%, and 1 wt.\,\%, respectively, and one close to c$^*$ with $c_p=$ 0.3 wt.\,\%. In table~\ref{tab:tab1}, we summarize the key material properties for the family of polymeric fluids used in the study. These polymeric solutions are all shear-thinning to various extents, with their viscoelasticity decreasing at lower concentrations, as shown in Fig.~\ref{fig:fig1-flowcurve} and Table~\ref{tab:tab2}.

The inelastic Cross model, $\eta(\dot{\gamma})=\eta_\infty+(\eta_0-\eta_\infty)/[1+(\lambda\dot{\gamma})^n]$ does an excellent job at describing the viscosity as a function of shear rate for all solutions (for best-fit parameter values see Fig.~\ref{fig:fig1-flowcurve}a and Table~\ref{tab:tab2}). Here, $\eta_0$ and $\eta_\infty$ are viscosities in the limit of zero and infinite shear rate, respectively, $\lambda^{-1}$ is a measure of the characteristic shear rate for the onset of shear thinning, and $n$ is the shear thinning index. 
The extent of shear-thinning in these viscoelastic solutions can be quantified in several ways. One direct metric is the ratio $\eta(\dot{\gamma})/\eta_0$. A more useful ratio that also provides consistency with earlier analysis, as we show later, is the dimensionless function $\beta_P(\dot{\gamma}) \equiv \eta_P(\dot{\gamma})/\eta(\dot{\gamma})=[\eta(\dot{\gamma})-\eta_{\infty}]/\eta(\dot{\gamma})$, which quantifies the relative polymer contribution to the total (rate-dependent) solution viscosity at a given shear rate.

\begin{table}
  \begin{center}
\def~{\hphantom{0}}
  \begin{tabular}{cccccc}
        $c_P$ (wt.\,\%) & $\rho$ (kg/m$^3$) & $\Gamma$ (mN/m) & $\eta_s$ (Pa.s) & $[\eta]$ (dL/g) & $c^*$ (wt.\,\%)\\ [3pt]
    0.30 $-$ 3.00 & 873.1 & 29.7 & 0.018 & 3.69 & 0.23 \\
  \end{tabular}
  \caption{PIB polymer solution properties.}
  \label{tab:tab1}
  \end{center}
\end{table}

\begin{table}
 \begin{center}
\def~{\hphantom{0}}
  \begin{tabular}{ccccc}
      $c_P$   & $\eta_0$   &  $\eta_{\infty}$  & $\lambda$  & n \\
      (wt.\,\%) & (Pa.s)  & (Pa.s) & (s) &   \\[3pt]
       3.0   & \hspace{-3mm}40.7 & 0.68 & 6.19 & 0.72 \\
       2.0   & 5.28 & 0.25 & 1.20 & 0.62 \\
       1.0  & 0.50 & 0.11 & 0.12 & 0.71\\
       0.3   & 0.10 & 0.05 & 0.06 & 0.81\\
  \end{tabular}
  \caption{Best fit parameters for the inelastic Cross model for the stable, steady-state rate-dependent shear viscosity data (Figure 1a) of various PIB solutions used in this study.}
  \label{tab:tab2}
 \end{center}
\end{table}

Shear-thinning rheology also means that the first normal stress coefficient $\Psi_1 = N_1/\dot{\gamma}^2$ and the relaxation time of the solutions are no longer material constants but become rate-dependent functions denoted by $\Psi_1(\dot{\gamma})$ and $\tau_s(\dot{\gamma})$. From the functional form of the upper-convected derivative, the first normal stress difference can be related to the shear stress and the relaxation time of the fluid through the expression $N_1(\dot{\gamma}) \simeq 2\tau_s\dot{\gamma}\sigma$ \citep{bird1987dynamics}, which after a slight rearrangement gives $N_1(\dot{\gamma})/\sigma(\dot{\gamma}) \simeq 2\tau_s\dot{\gamma}$. This ratio of the first normal stress difference $N_1$ to the shear stress in the sheared fluid $\sigma$ is termed the stress ratio $S_R$ and is a direct quantitative measure of non-linear viscoelastic effects in these polymeric solutions (see Figure~\ref{fig:fig1-flowcurve}b). The importance of non-linear elastic effects at the onset of flow instability in each fluid is evident as the stress ratio $S_R=N_1/\sigma \gtrsim 10$. One can also eliminate $\dot{\gamma}$ by substituting $\dot{\gamma} = \sigma/\eta$ in the expression above, which gives $N_1 \approx 2(\tau_s/\eta)\sigma^2$. So, one can anticipate on theoretical grounds that $N_1$ will vary quadratically with $\sigma$. The quadratic dependence of the first normal stress difference $N_1$ on the shear stress is well known for polymer solutions \citep{lodge1987measurement, binding1990shear} and has recently been documented in viscoelastic emulsions as well \citep{kibbelaar2023towards}. This key observation that $N_1 \sim \sigma^2$ will be crucial later in enabling us to incorporate shear-thinning rheology effects in a unified critical instability criterion (cf. Sec.~\ref{sec:unifying}).

\begin{figure}
  \centerline{\includegraphics[width=0.6\textwidth]{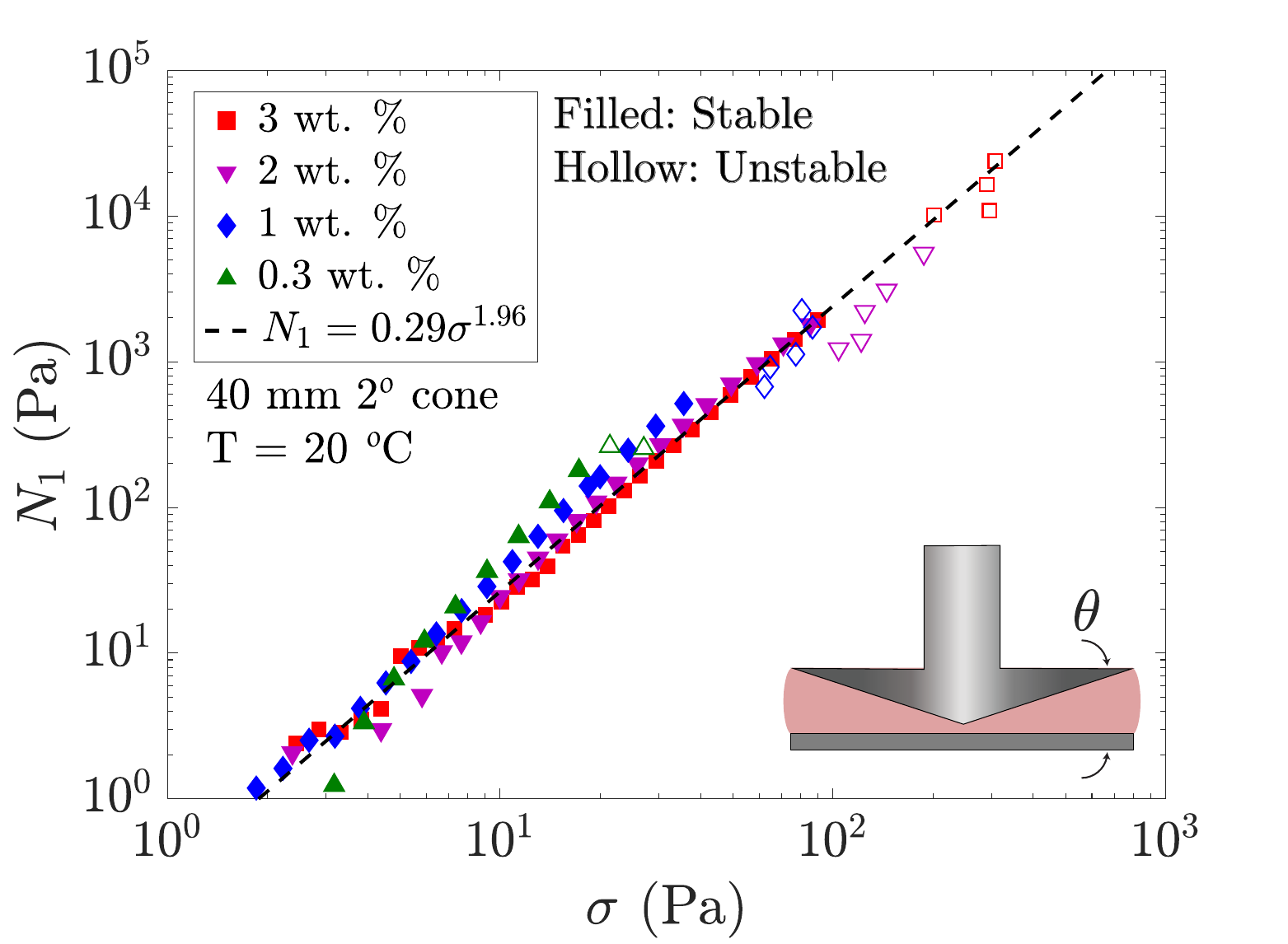}}
  \caption{The first normal stress difference $N_1$ plotted as a function of the shear stress $\sigma$ for all the PIB solutions used in this study in a 40 mm 2$^{\circ}$ geometry. The data lies almost on a single curve irrespective of the polymer concentration in solution with a power law best fit given by $N_1 \sim \sigma^{1.96}$, which is very close to the ideal quadratic dependence of $N_1$ on $\sigma$ expected theoretically.}
\label{fig:fig2-N1sigma}
\end{figure}

Furthermore, in semi-dilute polymer solutions, $\tau_s$ and $\eta$ are both typically power law functions of the polymer concentration $c_P$ \citep{heo2005scaling}. As a result, the modulus $G_c(c_P) \approx \eta(c_P)/\tau_s(c_P)$ only increases weakly with increasing the polymer concentration $c_P$. Hence, $N_1 \approx 2(\tau_s/\eta)\sigma^2 \approx \sigma^2/G_c(c_P)$ may be expected to only decrease weakly with the polymer concentration $c_P$ in the fluids utilized. We plot the first normal stress difference $N_1$ as a function of the shear stress $\sigma$ in Figure~\ref{fig:fig2-N1sigma} to test this prediction. We find that not only $N_1 \sim \sigma^2$, but this relationship also holds irrespective of the polymer concentration except when the flow becomes unsteady (hollow symbols). Thus, we can estimate the steady state values of $N_1(\dot{\gamma})$ from measurements of steady-state shear stress alone as:
\begin{equation}\label{eq:N1sigma}
    N_1(\dot{\gamma}) = A \, \sigma(\dot{\gamma})^2,
\end{equation}
where A is a constant coefficient obtained from the rheological measurements shown in Fig.~\ref{fig:fig2-N1sigma}. For the fluids utilized in this study, we find $A=0.29$ Pa$^{-1}$.

To incorporate rate-dependent rheological effects in the criterion for the onset of elastic instabilities (cf. Eq.~\ref{eq:eq1}), we use the more general definition of the Weissenberg number $Wi=N_1/2\sigma$ \citep{white1964dynamics}, which can be simplified using Eq.~\ref{eq:N1sigma} as:
\begin{equation}
    Wi \equiv \frac{N_1(\dot{\gamma})}{2\sigma(\dot{\gamma})} = \frac{A}{2}\sigma(\dot{\gamma}).
\end{equation}
Shear-thinning rheology also results in a rate-dependent characteristic shear relaxation time $\tau_s(\dot{\gamma})$, which enters Eq.~\ref{eq:eq1}. This relaxation time can be directly evaluated from experimental measurements of the first normal stress difference and the polymer contribution to the shear stress $\sigma_P=\eta_P(\dot{\gamma})\dot{\gamma}$. 
We can thus define $\tau_s(\dot{\gamma})=N_1/(2\eta_P(\dot{\gamma})\dot{\gamma}^2)$. After a slight rearrangement along with substituting $\sigma = \eta(\dot{\gamma})\dot{\gamma}$ and $\beta_P(\dot{\gamma}) = \eta_P(\dot{\gamma})/\eta(\dot{\gamma})$, we can thus write 
\begin{equation}\label{eq:taus}
    \tau_s(\dot{\gamma}) \equiv \frac{N_1(\dot{\gamma})}{2\eta_P(\dot{\gamma})\dot{\gamma}^2} =  \frac{1}{\beta_P({\dot{\gamma}})\dot{\gamma}} \, \frac{N_1(\dot{\gamma})}{2\sigma(\dot{\gamma})}. 
\end{equation}
This general expression for the rate-dependent characteristic relaxation time of a viscoelastic fluid can be combined with Eq.~\ref{eq:N1sigma} if it is validated experimentally for a given material system. For completeness, we note that this definition correctly reduces to the expression used in the literature (derived from the Oldroyd-B formulation) for the case of a non-shear-thinning Boger fluid; $N_1= 2\tau_s \sigma_P \dot{\gamma}=2 \tau_s \eta_P \dot{\gamma}^2$ where $\eta_P=\eta_0-\eta_s$, $\lim_{\dot{\gamma} \to 0}\Psi_{1} = \Psi_{1,0} = 2 \tau_s \eta_P$, and $\lim_{\dot{\gamma} \to 0}\tau_s(\dot{\gamma}) = \tau_s$. The rate-dependent shear relaxation time $\tau_s(\dot{\gamma})$ defined in Eq.~\ref{eq:taus} can also be used (if desired) to define a rate-dependent Deborah number $De=\tau_s(\dot{\gamma})\Omega$, which incorporates shear-thinning in the relaxation time. In addition, the normal stress difference ratio $\psi = -\Psi_{2,0}/\Psi_{1,0}$ for the PIB solutions utilized in this study fall in the range $0.205 - 0.243$ as measured from rod-climbing rheometry \citep{more2023rod}. Here $\lim_{\dot{\gamma} \to 0 }{\Psi_{2}} = \Psi_{2,0}$ with $\Psi_{2}$ being the second normal stress coefficient. We have included additional rheological characterization of the various PIB solutions were used in this study, including small amplitude oscillatory shear measurements in the online supplementary information.



\section{Results and Discussion}
As shown in Figure~\ref{fig:fig1-flowcurve}a, a sudden jump in the measured apparent  \textit{steady-state} shear viscosity is observed for all solutions beyond a sufficiently high shear rate when sheared in a cone and plate geometry. These measurements are not erroneous but are manifestations of the onset of time-dependent secondary motion beyond a critical (composition and geometry dependent) shear rate $\dot{\gamma}_c$. As a result, the stresses in the samples also suddenly increase due to large velocity disturbances. This is the origin of the so-called ``anti-thixotropic" transition \citep{jackson1984rheometrical}, observed and analyzed earlier for constant viscosity PIB Boger fluids \citep{mckinley1991observations, oztekin1994quantitative, schiamberg2006transitional}. In the present study, we analyze the same instability for viscoelastic fluids with pronounced shear-thinning, which inherently drives a transition of the instability mechanism from purely elastic to elastoinertial as we gradually decrease the polymer concentration. Finally, we use the following conventions for the governing rate-dependent dimensionless parameters: (1) a subscript $_0$ denotes dimensionless parameters determined using material function values in the zero-shear-rate limit, i.e., without incorporating rate-dependent viscosity effects, e.g., $Re_0=\rho \Omega R^2/\eta_0$, $Wi_0 = \tau_s \dot{\gamma}$, (2) the absence of any subscript denotes dimensionless parameters determined with the incorporation of rate-dependent viscosity effects, e.g., $Re = \rho \Omega R^2/\eta(\dot{\gamma})$, $Wi = \tau_s(\dot{\gamma}) \dot{\gamma}$, and (3) a subscript $_c$ denotes the critical condition $\dot{\gamma} \to \dot{\gamma}_c$, e.g., $Re_c=\rho \Omega_c R^2/\eta(\dot{\gamma}_c)$, $Wi_c = \tau_s(\dot{\gamma_c}) \dot{\gamma_c}$.

\subsection{Time-dependent instability, power spectra, and unsteady flow visualization}

\begin{figure}
  \centerline{\includegraphics[width=1\textwidth]{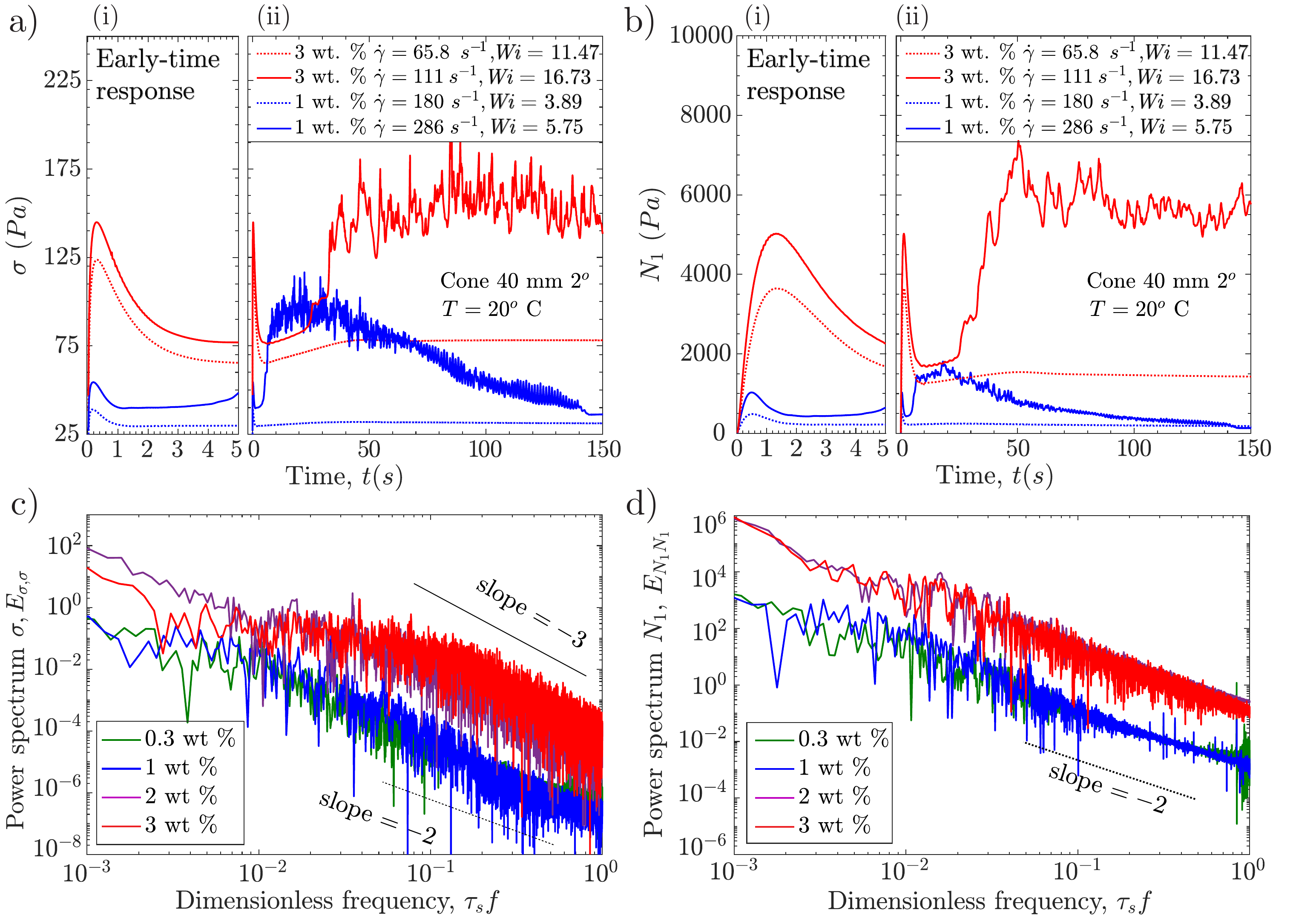}}
  \caption{Time-dependent evolution in a) shear stress and b) the first normal stress difference of the 3 wt.\,\% ($Re \lessapprox 1$) and the 1 wt.\,\% ($Re \gtrapprox 10$) samples in a peak hold experiment at shear rates below (dotted lines) and above (solid lines) the critical shear rate for the onset of instability. The corresponding values of $Wi = N_1/2\sigma=A \sigma/2$ are also given. Power spectra for c) shear stress and d) the first normal stress difference $N_1$ for the unstable flows shown in (a) and (b) are shown as a function of dimensionless frequency $\tau_s f$, and all four different polymer concentrations probed in this study. }
  
\label{fig:fig1-time+fft}
\end{figure}

The experimentally measured time-evolving shear stress $\sigma(t,\dot{\gamma})$ and first normal stress difference $N_1(t,\dot{\gamma})$ for a constant shear rate experiment in a 40 mm 2$^o$ cone and plate geometry are shown in Figure~\ref{fig:fig1-time+fft} for two different PIB fluids. The transient responses of $\sigma(t)$ and $N_1(t)$ are typical of viscoelastic fluids; at short times, a rapid stress increase with an overshoot (and initial quadratic growth in $N_1$) is observed. At longer times, constant steady-state values are observed only when the imposed shear rate is less than a critical value $\dot{\gamma}_c$. Beyond this point, both $\sigma(t,\dot{\gamma})$ and $N_1(t,\dot{\gamma})$ increase rapidly to a new fluctuating state similar to the earlier observations for Boger fluids \citep{mckinley1991observations}. Calculations of the apparent viscosity and first normal stress difference from averaging these enhanced time-dependent measurements result in the sudden jump observed in Figure~\ref{fig:fig1-flowcurve}. However, the time dependence of the unstable flow of the shear-thinning viscoelastic fluid has notable differences compared to a Boger fluid. 

As we reduce the PIB concentration, the viscosity of the solutions decreases, and inertial effects become increasingly important. The competition between the effects of shear thinning and the effects of inertia results in the onset of instability at a lower $Wi_c$ than the elastic shear-thinning case of the 3 wt.\,\% PIB solution. 
In addition, we observe a gradual decay in the mean amplitude of the stress fluctuations for the less concentrated PIB fluids due to slow irreversible sample ejection from the geometry edge as well as possible viscous heating effects \citep{calado2005transient}. As a result, the eventual steady-state value reached after the fluctuations in the $\sigma(t)$ and $N_1(t)$ response die out can be lower than the quasi-steady state value achieved after the initial transient is completed (at times $t\sim O(10)s$). Thus, reducing the PIB concentration gradually shifts the instability mechanism from purely elastic to elasto-inertial, and shear-thinning plays a crucial role in determining the critical conditions for the onset of instability. This transition in the instability mechanism becomes clear from the power spectra of the shear stress $\sigma$ (denoted $E_{\sigma\sigma}$) and the first normal stress difference $N_1$ (denoted $E_{N_1{N_1}}$) obtained by taking Fourier transforms of $\sigma(t)$ and $N_1(t)$ after the onset of time-dependent fluctuations. The corresponding power spectra for the fluid response (at $\dot{\gamma}>\dot{\gamma}_c)$ are presented in Figure~\ref{fig:fig1-time+fft}c and \ref{fig:fig1-time+fft}d, respectively. The slope of the power spectra progressively changes from $-3$ to $-2$ for $E_{\sigma\sigma}$ but remains unchanged as we gradually reduce the PIB concentration below 3 wt.\,\%. Furthermore, when plotted as a function of an appropriate dimensionless frequency $\tau_s(\dot{\gamma}) f$, the measured spectra fall onto two distinct curves: one for the more concentrated solutions (3 and 2 wt. \,\%) and a separate curve for the less concentrated solutions (1 and 0.3\,wt \%). This change in $E_{\sigma\sigma}$ appears to be a distinctive feature of a transition in the underlying instability mechanism \citep{steinberg2022new}.

\begin{figure}
  \centerline{\includegraphics[width=0.9\textwidth]{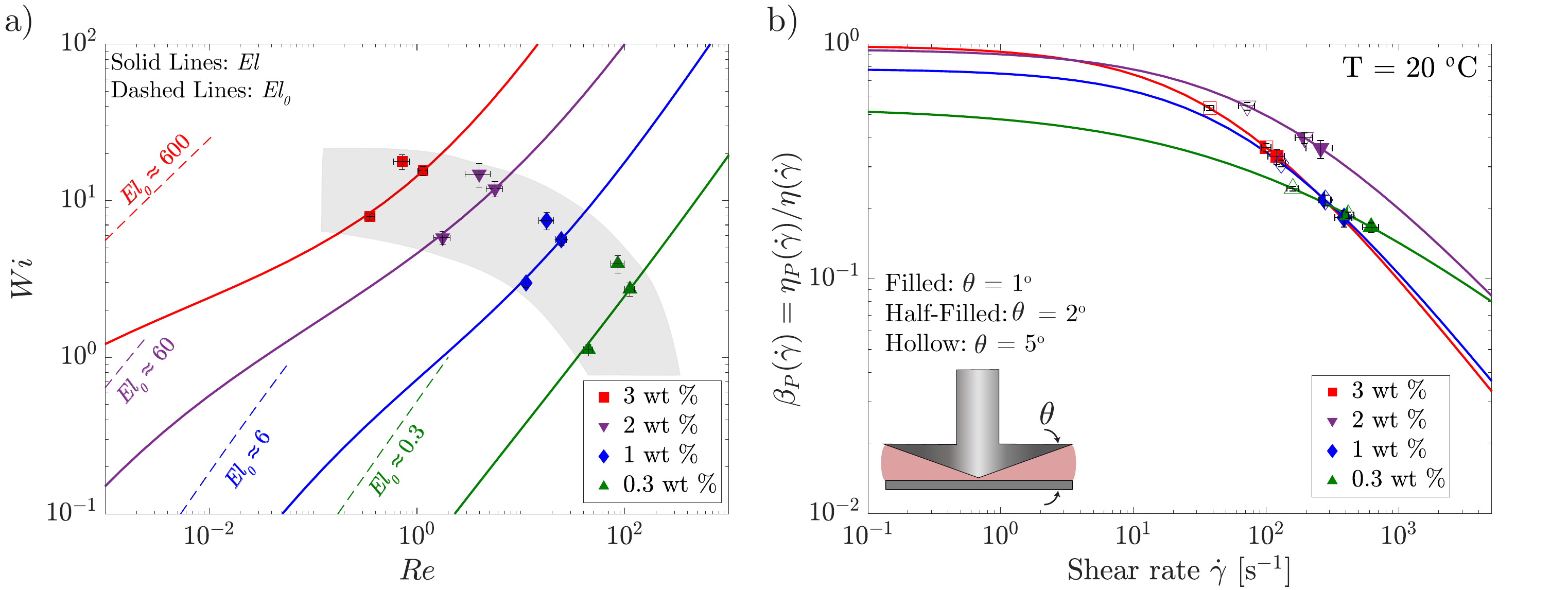}}

  \caption{Snapshot of the unsteady torsional flow between a cone-and-plate with a radius of 20 mm and 2$^{\circ}$ cone angle visualized from below. a-i) 3 wt. \% PIB solution rotating at $\Omega=3.5$ rad/s (Re = 1.15, Wi = 31.02). b-i) 1 wt. \% PIB solution at $\Omega=15$ rad/s (Re = 39.8, Wi=16.41). The instability starts at the outer rim and propagates radially inward. a-ii) Kymograph showing the temporal evolution of the flow along the rectangular strip marked by the dashed lines in (a-i). b-ii) Kymograph showing the time evolution of the flow in the rectangular region marked by the dashed lines in (b-i). The amplitude and the radial extent of the perturbations in the lower concentrated polymer solutions decrease with increasing fluid inertia.}
  
\label{fig:fig3-viz}
\end{figure}

Figure~\ref{fig:fig3-viz} shows instantaneous visualizations of the unsteady flow after the onset of instability in a 40 mm 2$^\circ$ cone geometry, and the corresponding space-time diagrams (kymographs) depict the evolution of the unsteady flow with time. The instability originates at the edge of the conical fixture and spirals toward the center (as can be seen from the fine dark streak spiraling inwards). As the perturbations propagate towards the center of the cone, they dissipate, which can be clearly seen in the kymographs; the undulating fluctuations are most intense near the edge but become dimmer towards the center of the cone. The progressive ejection of material from the gap due to time-dependent fluctuations is evident at long times.

\subsection{Hysteresis and determination of critical conditions for onset of instability}

Hysteresis has been observed for Boger fluids in the past where a sudden jump down in shear rate from a value larger than the critical value (after the induction of a time-dependent flow state) to a shear rate lower than the critical value that had initially been determined in a step ramp-up protocol resulting in the onset of time-dependent flow \citep{mckinley1991observations}. These observations are consistent with a subcritical Hopf bifurcation; however, directly accessing the hysteretic state and assessing its extent is easier if the control variable is switched to be the imposed stress. In the present work, we perform both controlled stress sweeps as well as stepped ramp-up in imposed shear rate measurements.

\begin{figure}
  \centerline{\includegraphics[width=0.6\textwidth]{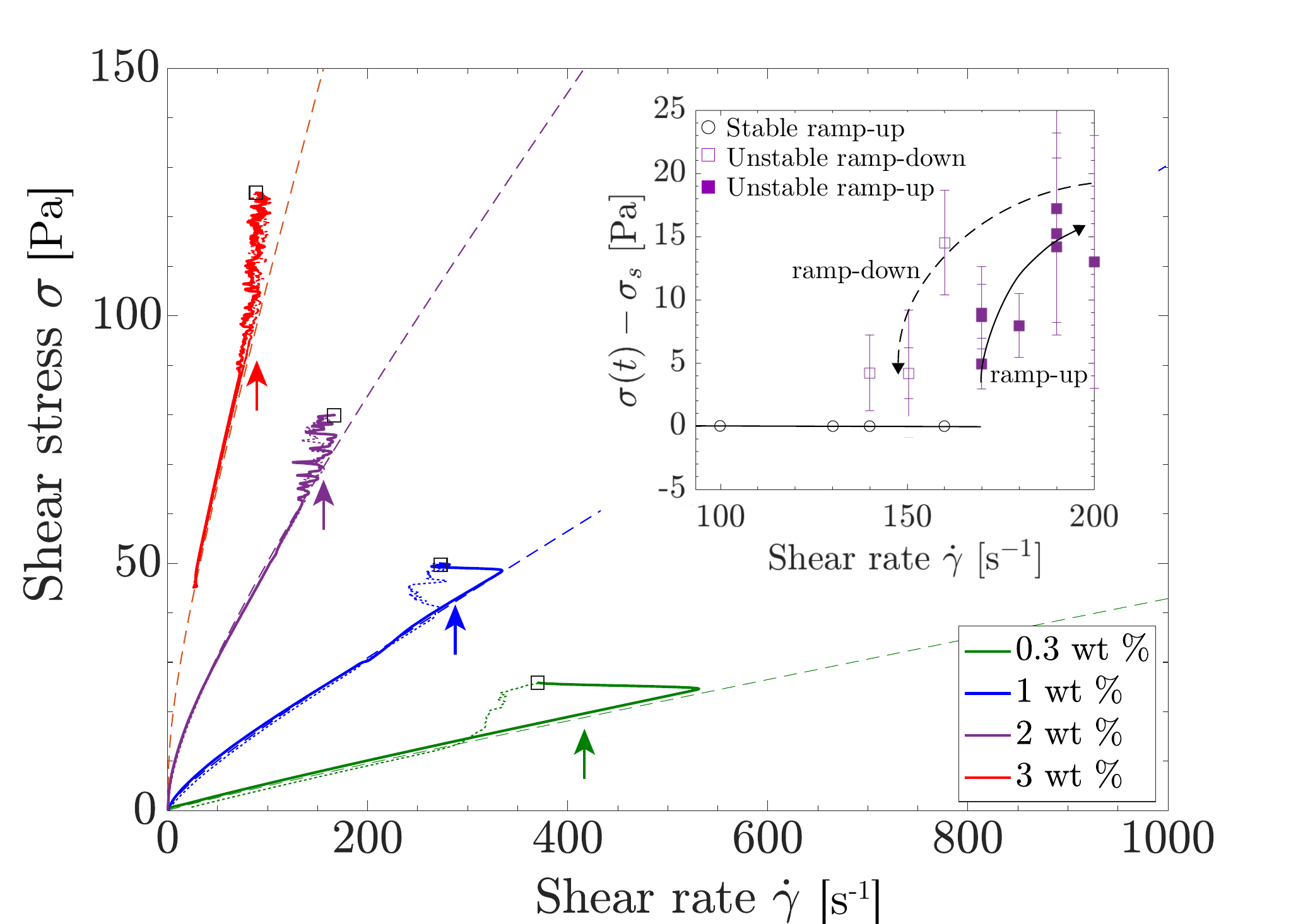}}
  
  \caption{Stress vs. shear-rate data for various PIB solutions used in this study for a continuous stress ramp-up experiment in a 40 mm 2$^{\circ}$ cone and plate geometry. In the continuous stress ramp-up experiment, fluids were subjected to a linearly increasing stress given by $\sigma(t) = \sigma_0 \left( 1+t/5\tau_s \right)$ (solid lines) and until maximum stress (indicated by hollow black square symbols for each fluid) in the unstable time-dependent flow region. The stress was reduced linearly from its maximum value (dotted lines) at the same rate as the ramp-up, and the instability was monitored by observing the resulting shear rate $\dot{\gamma}(t)$. Hysteresis in the critical shear rate can be inferred from whether the curve returns to its base state (shown by dashed lines, which is the Cross model fitted to the steady-flow curve data) or not. We observe that all the fluids undergo a hysteretic transition to an unsteady time-fluctuating state. The critical shear rate values measured from the stepped-up stress growth protocol are also shown as vertical arrows. The inset shows a hysteretic bifurcation curve in stepped shear rate ramp-up and down experiments for the 2 wt.\,\% fluid.} 
  
\label{fig:hysteresis}
\end{figure}

The critical values we report for the onset of instability, i.e., $\dot{\gamma}_c$, are determined by loading a fresh fluid sample and then performing a series of step increases in the shear rate. At each new shear rate, the evolution in the shear stress and the normal stress difference is followed for a long time ($\approx 30$ min) to observe the possible onset of a time-dependent unstable flow at a constant applied shear rate (see Fig.~\ref{fig:fig1-time+fft}a). The lowest shear rate at which such a transition was observed is then determined to be the critical shear rate. These critical values are indicated by the arrows in Fig.~\ref{fig:hysteresis}.  We call this a ``stepped shear rate ramp-up'' protocol. However, the transition might be hysteretic in the sense that if one has to “step down” from an unstable flow rate, the flow might only become stable at a shear rate lower than the critical value obtained in a “stepped ramp-up” experiment. To explore the presence and extent of flow hysteresis, we perform a continuous slow ramp-up and subsequently down in the shear stress (i.e., a saw-tooth stress profile) using the same geometry on a stress-controlled rheometer. In this protocol, we continuously increase the shear stress imposed on the sample (linearly in time) until the flow becomes unstable and then slowly and continuously reduce the applied stress. By “slow,” we mean that the rate of increasing the applied stress is much slower than the stress relaxation time for the fluids; here, we use a ramp rate of $1/5\tau_s$. In the absence of flow instability, such a protocol should trace out the steady flow curve predicted by the fit of the Cross-model for each fluid (shown by the dashed lines in Fig.~\ref{fig:hysteresis}). Hysteresis in the apparent flow curve of shear stress vs. shear rate can thus be directly confirmed by departures from this flow curve, as shown in Fig.~\ref{fig:hysteresis}.

It is evident from Fig.~\ref{fig:hysteresis} that the presence of strong viscoelastic shear-thinning reduces the magnitude of the flow hysteresis observed in the purely elastic case (for the 2 wt.\,\%, 3wt.\,\% fluids); however, careful measurements using the stepped shear rate ramp-up and ramp-down protocols and subtraction of the steady state flow stress $\sigma_s$ reveal that a small amount of hysteresis is still present (e.g., see the inset for the 2 wt.\,\% fluid). In the case of the inertio-elastic instability observed for the 0.3 and 1 wt.\,\% fluids, it is clear that the extent of flow hysteresis is very pronounced (with $\Delta \dot{\gamma}_c = \dot{\gamma}_c^{up} - \dot{\gamma}_c^{down} \approx 100\, s^{-1}$).

\subsection{Mapping the instability transition in the $Wi-Re$ plane}\label{sec:sec3p2}

\begin{figure}
  \centerline{\includegraphics[width=\textwidth]{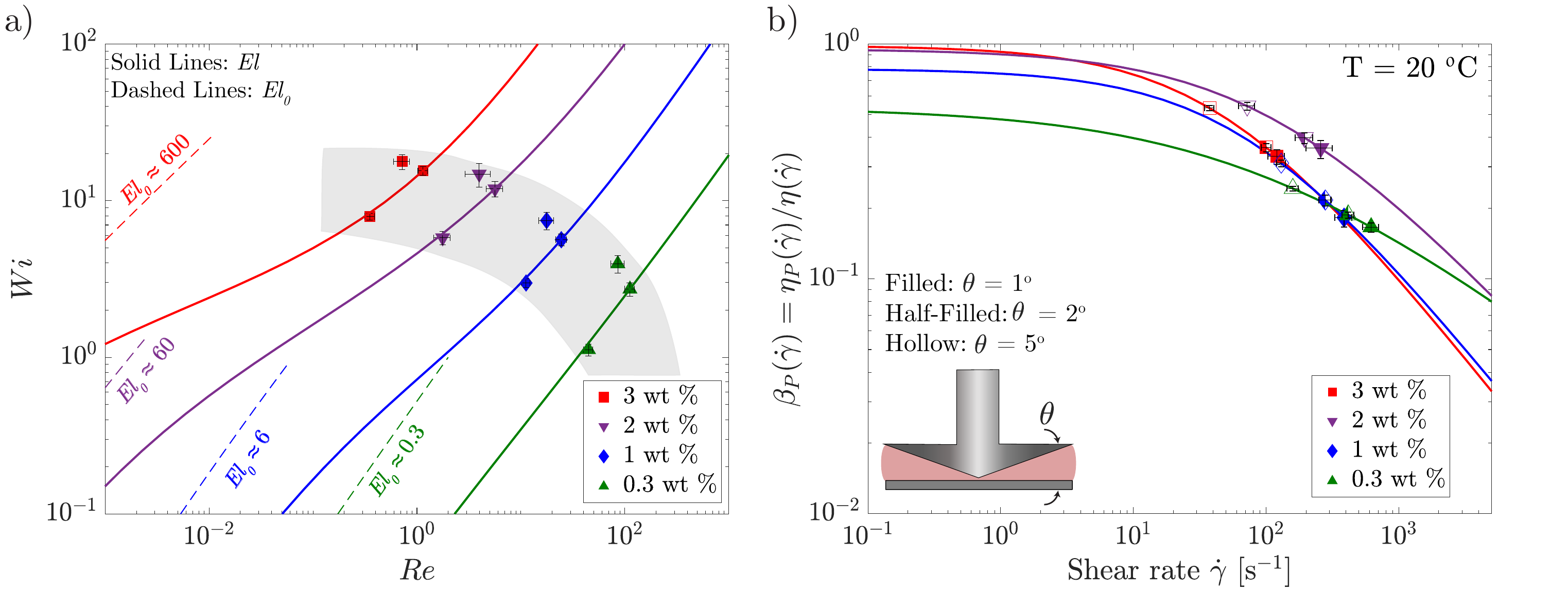}}
  \caption{a) Critical state diagram for the onset of complex instability for three different geometries and four fluid compositions projected onto the $Wi$–$Re$ plane. The samples tested span approximately four orders of magnitudes in elasticity number, and the instability mechanism shifts from purely elastic to elasto-inertial as we reduce the polymer concentration. Shear-thinning shifts the onset of instability to a higher $Re_c$ at a fixed $Wi$ as demonstrated by the two different curves representing the elasticity numbers defined by $El = Wi/Re$. The solid curve $El = Wi/Re$ incorporates the shear-thinning effect. The dashed lines are the zero shear-rate elasticity number $El_0=Wi_0/Re$ without considering shear-thinning in the viscosity. This effect becomes even clearer when we consider b) the shear-thinning parameter $\beta_P(\dot{\gamma})$ as a function of the shear rate and the critical shear rates for the onset of time-dependent flows for the different PIB solutions in different geometries. Symbols are experimentally determined critical values of $\dot{\gamma}$ and the associated flow parameters $Wi_c$ and $Re_c$.}
\label{fig:fig4-state}
\end{figure}

The changes in the magnitude of the stress fluctuations and the power law slope of the power spectra suggest a transition in the underlying mechanism behind the onset of instability as we change the polymer concentration. Experiments with a range of cone-and-plate geometries ranging from $\{R,\theta\}=\{ 2\textrm{ cm}, 1^\circ\}$ to $\{ 1.25\textrm{ cm}, 5^\circ\}$ also show that the critical conditions vary systematically with changes in the geometric parameter $1/\theta$. These changes can be understood more quantitatively by projecting the critical conditions for the onset of instability onto the $Wi-Re$ plane \citep{datta2022perspectives} as illustrated in Figure~\ref{fig:fig4-state}a. The transition from purely elastic to elasto-inertial can be clearly demonstrated by considering the evolution in the elasticity number $El=Wi/Re$ with shear-thinning effects (solid lines). The elasticity number quantifies the relative magnitude of viscoelastic and inertial effects in the $Wi-Re$ state diagram. The trajectory followed by highly elastic fluids with a constant viscosity $\eta_0$ (i.e., Boger fluids) is represented in Figure~\ref{fig:fig4-state}a by dashed lines with constant slopes ($El_0=Wi_0/Re$). By including the effect of shear-thinning through the rate-dependent viscosity of the PIB solutions, we can conclude that shear-thinning shifts the critical Reynolds number $Re_c$ for the onset of instability to substantially higher value (at a given Weissenberg number). The value of the elasticity number at the onset of instability $El_c$ reduces by four orders of magnitude (from $\approx$ 800 to 0.02) as we reduce the concentration of the PIB solution by one order of magnitude ( from 3 wt.\,\% to 0.3 wt.\,\%), signifying the rapid rise in inertial effects compared to viscoelastic effects.

At low $Re$, shear-thinning increases the stability of the torsional shear flow of viscoelastic fluids against the onset of purely elastic instabilities. For example, the Boger fluid used by \cite{schiamberg2006transitional} exhibited purely elastic instability beyond $Wi_c \approx O(1)$ compared to the values of $Wi_c \sim O(10)$ required for an elastic but strongly shear-thinning PIB solution. However, reducing the concentration of the PIB solutions also decreases the extent of shear-thinning, i.e., the magnitude of changes in $\beta_P$ are reduced, and the onset of instability shifts to $Wi_c \sim O(1)$. Thus, reducing the extent of shear-thinning (or maintaining $\beta_P$ closer to unity) drives the earlier onset of elastic instability when represented in terms of $De$ or $Wi$.

The gradual increase in inertial effects is another crucial factor contributing to shifts in the onset conditions of the instability with a reduction in the PIB concentration. It is observed that the presence of elasticity makes the flow unstable at much lower Reynolds numbers ($Re_c \lessapprox 100$) compared to the critical value of $\rho \Omega_c R^2/\eta$ required for the onset of purely inertial turbulence in cone-plate flows of Newtonian fluids. Finally, the shaded region, which is included solely as a guide for the eye, hints at common trends and the possible collapse of all the critical conditions onto a single master curve. In the following section, we seek a unified critical condition spanning purely elastic to elastoinertial regimes for a range of polymer concentrations and flow geometries.  

\subsection{Bridging elasticity and inertia: A unified instability criterion for the onset of instability in torsional flows of shear-thinning viscoelastic fluids}\label{sec:unifying}

\begin{figure}
\centering
  \centerline{\includegraphics[width=\textwidth]{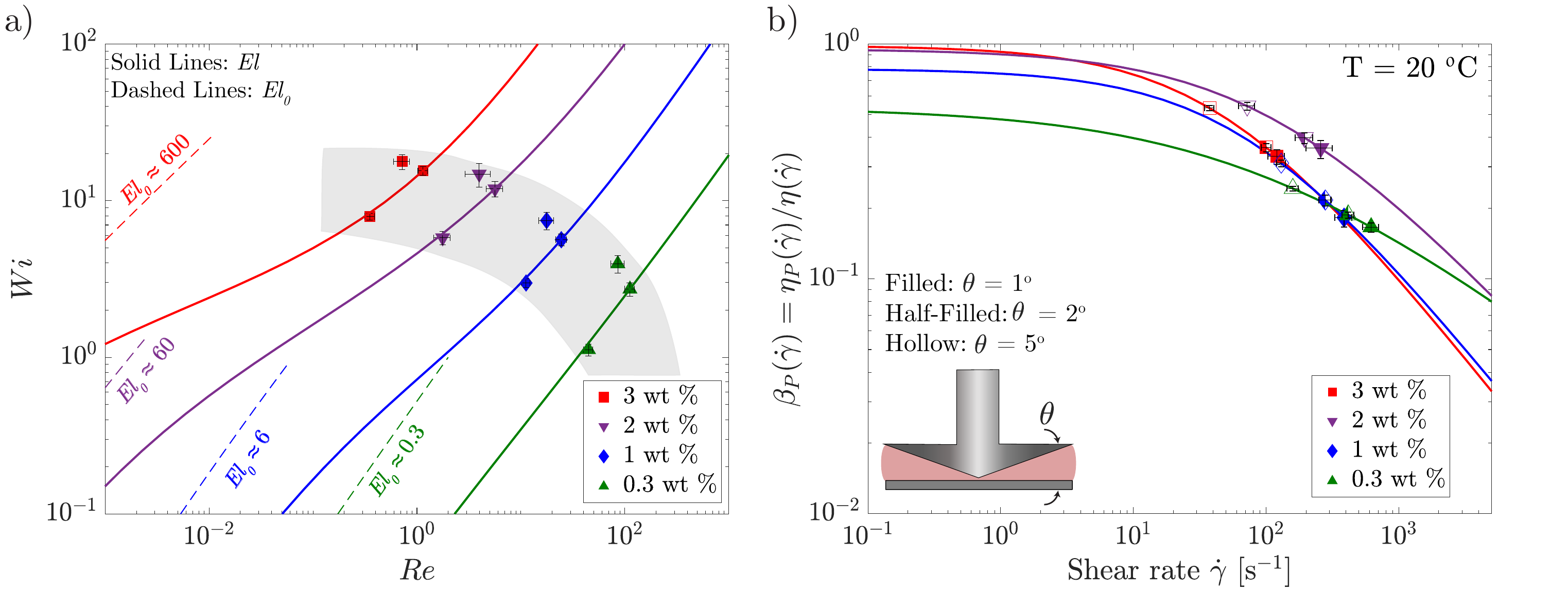}}

  \caption{a) The predictions of the upgraded purely elastic instability criterion incorporating the effects of shear-thinning in fluid rheology Eq.~\ref{eq:condST} using the values of $\beta_P({\dot{\gamma_c}})$ shown in Figure~\ref{fig:fig4-state}b for four different fluids used in four different geometries. Boger fluid results are also presented for comparison. It can be concluded that the data lies on different stability curves (dotted lines to guide the eye), which are shifted due to the presence of inertia. Hence, a single elasto-inertial stability curve should incorporate inertial effects. b) Updated state diagram obtained by scaling the axes to incorporate the effects of fluid elasticity, inertia, shear-thinning, and geometry using Eq.~\ref{eq:eq6}. This state diagram describes the onset of elasticity over a wide range of $Re$ and $Wi$ spanning four decades in the elasticity number $El$ and four different shear-thinning viscoelastic fluids. Dashed lines show $68$ \% confidence interval corresponding to one standard deviation interval. Thus, the scaling presented in Eq.~\ref{eq:eq6} can be used to understand changes in the critical conditions in purely elastic and elasto-inertial instabilities.}

\label{fig:fig5-collapse}
\end{figure}

The governing role of shear-thinning in stabilizing the flow of the PIB solutions can be seen from the magnitudes of the decrease in $\beta_P(\dot{\gamma})$ shown in Fig.~\ref{fig:fig4-state}b with shear rate and/or decreasing PIB concentration. In addition, Eq.~\ref{eq:taus} shows that the shear relaxation time is also a shear-thinning function of the shear rate, and this plays an important role in determining the onset of instability. These non-linear effects of shear thinning can be incorporated heuristically in the critical conditions for the onset of a purely elastic instability by incorporating the rate-dependence of the relaxation time $\tau_s(\dot{\gamma})$ from Eq.~\ref{eq:taus} into Eq.~\ref{eq:eq2}:
\begin{equation}\label{eq:eq3}
    \frac{\tau_s \, {U}}{\mathcal{R}} \, \frac{N_1}{\sigma} \equiv \frac{\tau_s(\dot{\gamma}) \, {U}}{\mathcal{R}} \, \frac{N_1}{\sigma} = \frac{{U}}{\mathcal{R}} \frac{2}{\beta_P(\dot{\gamma}) \dot{\gamma}} \left( \frac{N_1}{2\sigma} \right)^2 \geq \, M_c^2. 
\end{equation}
In a cone and plate geometry with cone angle $\theta$ and radius $\mathcal{R}=R$ rotating with a rate $\Omega$ we get the characteristic velocity $U=\Omega \, R$ and shear rate $\dot{\gamma}=\Omega/\theta$. Using these characteristic values and the general definition of the Weissenberg number $Wi = N_1/2\sigma$, a further simplification of Eq.~\ref{eq:eq3} can be obtained for a cone and plate geometry:
\begin{equation}\label{eq:condST}
    \frac{2\theta}{\beta_P(\dot{\gamma})} \, \left( \frac{N_1}{2\sigma} \right)^2 = \frac{2\theta}{\beta_P(\dot{\gamma})} \, Wi^2  \geq \, M_c^2.
\end{equation}
Thus, Eq.~\ref{eq:condST} gives a critical condition for the onset of purely elastic instability incorporating the effects of shear thinning, viscoelasticity, and changes in flow geometry. We note that the critical condition of Eq.~\ref{eq:eq2} for the onset of purely elastic instabilities in constant viscosity Boger fluids \citep{mckinley1996rheological} can be recovered from Eq.~\ref{eq:condST} by substituting the well-known Oldroyd-B results for $N_1 = 2\eta_P\tau_s\dot{\gamma}^2$ and $\sigma=\eta_0\dot{\gamma}$. Curves of neutral stability consistent with Eq.~\ref{eq:condST} can also be drawn in the $Wi - \beta_P(\dot{\gamma})/\theta$ plane, where the effects of shear thinning can be incorporated by shifting the geometric parameter $1/\theta$ by an amount depending on the shear-thinning parameter $\beta_P(\dot{\gamma})$ \citep{oztekin1994quantitative}. These curves of neutral stability (dotted lines) incorporating shear-thinning and geometry effects in the purely elastic critical instability criterion along with the experimentally measured critical conditions are presented in Fig.~\ref{fig:fig5-collapse}a using the values of $\beta_P({\dot{\gamma_c}})$ shown in Figure~\ref{fig:fig4-state}b. We can immediately conclude that the critical conditions shown in Figure~\ref{fig:fig4-state}a would only be shifted vertically by this scaling. This modified critical stability condition still does not incorporate the effects of fluid inertia.

A unified critical condition must ultimately be deduced by performing a rigorous linear stability analysis of the elasto-inertial flow problem. In the absence of any such existing analysis, our empirical measurements can be harnessed to guide the formulation of a unified critical condition. Purely elastic instability arises due to non-linearities in the fluid constitutive equations, while purely inertial Newtonian turbulence arises from the non-linearities in the advective term of the equation of motion. The distinct origins of the two sources of non-linearity suggest that from the viewpoint of infinitesimal perturbations, the two destabilizing terms may be coupled together so that instability ensues when some combination of elastic effects and inertial effects becomes larger than a threshold or a critical value. 

The effects of non-linear elasticity and shear thinning on the steady base torsional shearing flows are already included in Eq.~\ref{eq:eq3}; this \textit{purely elastic} criterion just needs to be augmented by the appropriate non-linear term to incorporate the effects of fluid inertia in determining the onset of instability. In addition, this unified criterion should 1) recover the critical instability condition for the onset of purely elastic instability in the absence of inertia, 2) recover the critical instability condition for the onset of purely inertial instability in the absence of elasticity, and 3) predict a smooth transition between the two asymptotic regimes as observed in Fig.~\ref{fig:fig4-state}a.

Detailed linear stability analyses \citep{joo1994observations, oztekin1994quantitative} consider infinitesimal perturbations to the base flow velocity and stress variables $(\mathbf{v}^0, \boldsymbol{\sigma}^0)$ as well as the corresponding velocity and stress gradients $(\mathbf{\nabla} \mathbf{v}^0, \mathbf{\nabla}\boldsymbol{\sigma}^0)$ present in the specific flow field of interest. In the inertialess limit, the perturbed velocity and stress fields can be written in the dimensionless form $\mathbf{v} = \mathbf{v}^0 + Wi \mathbf{v}' + O(Wi^2)$ and $\boldsymbol{\sigma} = \boldsymbol{\sigma}^0 + Wi \boldsymbol{\sigma}' + O(Wi^2)$ (where the stress tensor has been non-dimensionalized with the characteristic viscous stress $\sim \eta_0 U/\mathcal{R}$). Substituting these forms into the governing equations and collecting terms at the first order results in a complex eigenvalue problem involving terms of the form $Wi(\mathbf{\nabla} \mathbf{v}^{0} \cdot \boldsymbol{\sigma}')$, $Wi (\mathbf{v}^0 \cdot \mathbf{\nabla}\boldsymbol{\sigma}')$, etc. (see \cite{joo1994observations, shaqfeh1996purely} for details). From a scaling viewpoint, the resulting stability criterion is always of the form of Eq.~\ref{eq:eq1}, or equivalently,
\begin{equation}\label{eq:elastic}
    \sqrt{\tau_s \frac{U}{\mathcal{R}} \, \frac{N_1}{\sigma}} \geq M_c. 
\end{equation}

Similarly, for the onset of Newtonian curved streamline instabilities, the perturbations can be written in the form $\mathbf{v} = \mathbf{v}^0 + Re \mathbf{v}' + O(Re^2)$ and the eigenvalue problem arises from coupling between the advective terms of the form $Re \mathbf{v}^0 \cdot \mathbf{\nabla} \mathbf{v}'$, $Re \mathbf{v}' \cdot \mathbf{\nabla} \mathbf{v}^0$, etc. From a scaling viewpoint, the resulting relevant dimensionless group that arises is the G\"ortler number $G = Re (\delta/\mathcal{R})^{3/2}$ \citep{saric1994gortler}, which couples streamline curvature parameterized by a characteristic radius of curvature $\mathcal{R}$, inertial effects parameterized by a Reynolds number, and a small parameter $\epsilon = \delta/\mathcal{R}$ parameterized by the ratio of boundary layer thickness $\delta$ to the radius of curvature $\mathcal{R}$. In a Taylor-Couette flow, on the other hand, the relevant dimensionless group that arises is the Taylor number $Ta=Re^2\epsilon^3$, which also couples inertial effects parameterized by Reynolds number and a small parameter $\epsilon = h/\mathcal{R}$ where $h$ is the gap between two concentric cylinders. The dimensionless grouping on the left-hand side of Eq.~\ref{eq:elastic} can be viewed as a viscoelastic G\"ortler number \citep{pakdel1996elastic}. These arguments all suggest that inertial effects can be parameterized by combining the Reynolds number and a flow geometry-dependent small parameter. So, to quantify inertial non-linearities, we use the dimensionless ratio of the dynamic pressure $\rho{U}^2$ compared to the shear stress $\sigma$, i.e., 
\begin{equation}\label{eq:inertial}
    \frac{\rho U^2}{\sigma} \geq C^2 \quad \textrm{     or equivalently    } \sqrt{\frac{\rho U^2}{\sigma}} \geq  C,
\end{equation}
where $C$ is some critical value of this ratio. We note that this ratio is the same functional form as the dimensionless ratio $N_1/\sigma$, which determines the purely elastic instability and, as \cite{renardy2000current} has noted, in strongly non-linear elastic flows that streamline tension, gives rise to a modified form of the inviscid Euler or Bernoulli equation. This scaling in Eq.~\ref{eq:inertial} automatically gives rise to a combination of Reynolds number and a small parameter to quantify the inertial effects, similar to G\"ortler and Taylor numbers, as will be evident in the discussion below. 

A full inertioelastic linear stability analysis must consider perturbations arising from both elastic effects and inertial effects as well as cross-coupling terms that scale as $Re\,Wi\,\mathbf{v}'$. The resulting eigenvalue problem will be extremely complex, but from a dimensional viewpoint, it must embody the two limits, purely elastic and purely inertial, discussed above, as well as relevant cross-terms. The simplest possible functional form that captures these limits as well as cross-coupling between the elastic stresses and inertial perturbations is a linear combination of the two dimensionless groupings in Eq.~\ref{eq:elastic} and Eq.~\ref{eq:inertial}. This linear combination can be written in the generic form
\begin{equation}\label{eq:eq5}
    \sqrt{\frac{\tau_s \, {U}}{\mathcal{R}} \, \frac{N_1}{\sigma}} + \alpha \sqrt{ \frac{\rho{U^2}}{\sigma}} \geq \tilde{M}_c \quad \textrm{     or equivalently    } \left[ \sqrt{\frac{\tau_s \, {U}}{\mathcal{R}} \, \frac{N_1}{\sigma}} + \alpha \sqrt{ \frac{\rho{U^2}}{\sigma}} \right]^2 \geq \tilde{M}_c^2, 
\end{equation}
where $\alpha$ is the dimensionless weighting parameter, and $\tilde{M}_c$ is the critical disturbance magnitude (as modified by the inertial effects). Expanding the square in the second equality captures the critical criterion. Further simplification of Eq.~\ref{eq:eq5} in a cone and plate geometry can be obtained by using Eq.~\ref{eq:condST} and substituting the characteristic velocity $U=\Omega{R}$ and $\dot{\gamma} = \Omega/\theta$ so that at the critical condition $\dot{\gamma} \to \dot{\gamma}_c$ we require
\begin{equation}\label{eq:eq6}
    \sqrt{\frac{2\theta}{\beta_P(\dot{\gamma})} \,Wi^2} + \alpha \, \sqrt{Re \, \theta} \geq \tilde{M}_c \quad \textrm{     or equivalently    } \quad \theta \left[ \sqrt{\frac{2Wi^2}{\beta_P(\dot{\gamma})}} + \alpha \sqrt{Re} \right]^2 \geq M_c^2,
\end{equation}
where a rate-dependent Reynolds number $Re=\rho\Omega{R^2}/\eta(\dot{\gamma})$ and also the (small) geometry parameter $\theta$ appears naturally in the inertial term. This new generalized criterion in Eq.~\ref{eq:eq6} asymptotically recovers the purely elastic instability condition (Eq.~\ref{eq:condST}) when inertia is negligible, i.e., $Re \to 0$, and also predicts the onset of secondary motion due to inertial effects in Newtonian fluids in a cone and plate geometry when the product $Re\,\theta$ exceeds a critical value consistent with earlier perturbation analyses and experiments \citep{fewell1977secondary, sdougos1984secondary}.


In lieu of formal linear stability analysis, this unified critical condition can be validated empirically with experimental measurements. We use the result from Eq.~\ref{eq:N1sigma}, which indicates that $N_1$ varies quadratically with $\sigma$ to calculate the critical Weissenberg number $Wi_c = A\sigma(\dot{\gamma}_c)/2$ with $A=0.29$ obtained from the regression of the experimental measurements presented in Figure~\ref{fig:fig2-N1sigma}. If the condition suggested by Eq.~\ref{eq:eq6} is true, then plotting data for the onset of instability in the $Wi^2 \theta/\beta_P(\dot{\gamma})$ vs. $Re \theta$ phase space should lie on a single curve. We find that this is indeed true for a suitable value of the dimensionless weighting parameter $\alpha=2.29$ as shown in Figure~\ref{fig:fig5-collapse}b. The collapse of the data over three orders of magnitude in the elasticity number for four different viscoelastic fluids with varying shear-thinning strengths and for three different conical geometries strongly corroborates the functional form of the unified critical condition for the elasto-inertial instability given in Eq.~\ref{eq:eq6}. Linear regression to experimental measurements of the critical conditions gives $\tilde{M}_c \approx 5.46$.


A further check of the unified critical condition Eq.~\ref{eq:eq6} can be performed by checking if $\tilde{M}_c \approx 5.46$ obtained from our experimental measurements is consistent with critical conditions in the limiting cases of purely elastic instability of a Boger fluid and a purely inertial instability of a Newtonian fluid in a cone and plate geometry. We, therefore, project the dimensionless criterion represented by Eq.~\ref{eq:eq6} (with $\alpha \simeq 2.29$ and $\tilde{M}_c \simeq 5.46$) in the modified $Wi-Re$ plane to incorporate the effects of shear-thinning and flow geometry as shown in Fig.~\ref{fig:fig5-collapse}b. Our heuristic blending rule provides a good collapse of all the experimental data shown in Fig.~\ref{fig:fig4-state}a. A typical non-shear-thinning Boger fluid has $\beta_P({\dot{\gamma}}) \equiv \beta \simeq 0.4$ and $Re \to 0$, which recovers $Wi_c \approx 60$ consistent with previous experimental observations \citep{mckinley1991observations}. On the other hand, for a Newtonian fluid, Eq.~\ref{eq:eq6} predicts the onset of the unsteady secondary flow in a cone and plate geometry at $Re\,\theta \lesssim 5.68$, which is consistent with, but smaller than, the value deduced from early visualization experiments of \cite{sdougos1984secondary}. This scaling form can, in principle, be readily extended to other geometries by the appropriate determination of the radius of curvature of the streamlines, the characteristic flow velocity, and the identification of the appropriate characteristic shear rate \citep{mckinley1996rheological}. For example, in a parallel plate geometry, the geometric factor would be $H/R$, where $H$ is the gap height between the plates. 

In the spirit of Occam’s razor, we have proposed the simplest possible physically reasonable criterion (additively combining the destabilizing effects of elasticity and inertia). We hope that our results will serve as a motivator for a more detailed stability analysis that fully considers the cross-coupling of perturbation terms between the equation of motion and the nonlinear viscoelastic constitutive equation. 

\section{Conclusions}
We have investigated the onset of time-dependent instabilities in the torsional shearing flow between a cone and a plate with $1^{\circ} \leq \theta \leq 5^{\circ}$ for a range of viscoelastic shear-thinning fluids. The onset of instability depends on the coupling between the fluid elasticity, inertia, shear-thinning, and flow geometry. Specifically, shear-thinning in the fluid rheology stabilizes the unidirectional torsional base flow, while inertia destabilizes it. Furthermore, the systematic changes in the critical conditions and the change in the slope of the power spectra $E_{\sigma \sigma}$ suggests a transition in the instability mechanism from purely elastic to elasto-inertial as we reduce the polymer concentration (and the extent of shear-thinning). We show how to represent this multidimensional flow instability transition by constructing a state diagram of critical conditions in the $Wi-Re$ plane. Finally, we propose a unified criterion (Eq.~\ref{eq:eq5}) to predict the onset of elasto-inertial instability by augmenting the existing criterion for purely elastic instability to include shear-thinning and coupling to perturbations arising from inertial effects in the Cauchy momentum equation. We validate the form of this unified criterion empirically with our experimental measurements of the critical conditions for a range of viscoelastic fluids and different conical geometries. The data collapse over a wide range of conditions corroborates the final functional form of Eq.~\ref{eq:eq6} at least for a cone-and-plate geometry. Extending this unified criterion to other torsional flow geometries (like the parallel plate geometry and Taylor-Couette flow) using an appropriate G\"ortler number would provide further verification of this representation. \\

\footnotesize\textbf{Acknowledgments}: The authors would like to thank Lubrizol Inc. for financial support and for providing the components required to formulate the polymeric fluids used in this study. 

\footnotesize\textbf{Declaration of interests}: The authors report no conflict of interest.


\printbibliography 

\end{document}